\documentclass[12pt]{iopart}

\usepackage{iopams}

\usepackage{graphicx}
\usepackage{mathrsfs}
\usepackage{verbatim}
\usepackage{pifont} 

\usepackage{hyperref}

\newcommand{\be}{\begin{equation}}
\newcommand{\ee}{\end{equation}}
\newcommand{\bea}{\begin{eqnarray}}
\newcommand{\eea}{\end{eqnarray}}

\newcommand{\hbo}{\hbox to 1 true cm {\hfill } }

\newcommand{\ud}{\mathrm{d}}

\newcommand{\fa}{\cite{Faddeev:2008qc}}
\newcommand{\ma}{\cite{Masson:2010vx}}

\begin{document}

\title{Symmetry breaking, conformal geometry and gauge invariance.}

\author{Anton Ilderton$^1$, Martin Lavelle$^2$, David McMullan$^3$}

\address{$^1$Department of Physics, Ume\aa\ University, 901-87 Ume\aa, Sweden}
\address{$^{2,3}$Particle Theory Group, School of Mathematics, University of Plymouth, Plymouth PL4 8AA, UK}
\ead{anton.ilderton@physics.umu.se, martin.lavelle@plymouth.ac.uk, david.mcmullan@plymouth.ac.uk}

\begin{abstract}
When the electroweak action is rewritten in terms of SU(2) gauge invariant variables, the Higgs can be interpreted as a conformal metric factor. We show that asymptotic flatness of the metric is required to avoid a Gribov problem: without it, the new variables fail to be nonperturbatively gauge invariant. We also clarify the relations between this approach and unitary gauge fixing, and the existence of similar transformations in other gauge theories.
\end{abstract}


\section{Introduction}
A reinterpretation of the Weinberg--Salam model has recently appeared in the literature \cite{Chernodub:2008rz, Faddeev:2008qc, Masson:2010vx}, in which a change of variables is used to transform the action into one depending only on SU(2) {\it invariant} fields. In this way, the local SU(2) symmetry is factored out of the action. In addition, the approach of \cite{Chernodub:2008rz, Faddeev:2008qc} omits the usual Higgs symmetry breaking terms, and reinterprets the Higgs field as a dilaton. The action then describes a gravitational theory in which the electroweak fields interact in a locally conformally flat spacetime. In this picture, it is the condition of asymptotic flatness which is responsible for mass generation.

The transformation of \cite{Chernodub:2008rz, Faddeev:2008qc, Masson:2010vx} was actually considered some time ago in  \cite{Vlasov:1987vt, Lunev:1994ty} and \cite{Lavelle:1994rh},  where it was indeed used to construct gauge invariant variables, and in \cite{Langguth:1985dr, Montvay:1985nk, Hasenfratz:1987uc} where it was put to various uses on the lattice. (See \cite{Goldstone:1978he, Frohlich:1981yi, Farhi:1995pt, Gromov:2007he, Gromov:2010gf} for related and other approaches to gauge invariant variables.) In this paper we will bring together the various interpretations of the transformation, clarify its relation to gauge fixing and point out that the transformation potentially suffers from an ambiguity at the quantum level, which can be interpreted as a Gribov problem. From this we will obtain an important physical consequence of the asymptotic flatness condition: it is precisely this which allows us to sidestep the Gribov ambiguity in this situation, and so obtain nonperturbatively locally gauge invariant definitions of the electroweak fields, including the $Z$ and $W$ gauge bosons. This is in contrast to unbroken SU(3), where there is no gauge invariant description of the gluon, which consequently is not observable.

We begin in Sect.~\ref{review} with a brief review of the classical transformation to gauge invariant variables, the quantum contributions arising from the functional determinant for the transformation, and the interpretation of the Higgs as a conformal metric factor. In Sect.~\ref{gauge-sect} we explain the relations between this approach, unitary gauge fixing in the matter sector, and the definition of physical charges in gauge theories. In Sect.~\ref{check-sect} we show that a careful treatment of boundary conditions is necessary in order for the given transformation to make sense nonperturbatively. Here we also explore the connections between the scalar field's v.e.v.\ and the Gribov ambiguity. We also discuss the (non) existence of similar transformations in other gauge theories. The important details of the new interpretation also holds for the SU(2) Higgs model, and we work for the most part with this theory in order to simplify the presentation and clarify what is really going on. The extension to SU(2)$\times$U(1) is direct and will be described in Sect.~\ref{check-sect}. We conclude in Sect. \ref{Concs}.

\section{To gauge invariant variables}\label{review}

\subsection{The classical transformation}
We begin with the transformation to gauge invariant variables in the SU(2) Higgs model, which has the Lagrangian
\be\label{lag1}
	\mathcal{L} = |D_\mu\Phi |^2 - \frac{1}{4}F^a_{\mu\nu}F^{\mu\nu}_a + \mathcal{L}_{ssb}(|\Phi|)\;.
\ee
Here $\Phi=(\Phi_1,\Phi_2)$ is a doublet of complex scalars, with covariant derivative $D_\mu = \partial_\mu + g A_\mu$, and our $A_{\mu}$ are anti--hermitian. We consider two forms of the potential terms, 
\bea
\label{ssb1}	\mathcal{L}_{ssb}(|\Phi|) &=& 0\;,\\
\label{ssb2}	\mathcal{L}_{ssb}(|\Phi|) &=& -m^2 |\Phi|^2 -\lambda |\Phi|^4\;,
\eea
the second being the usual symmetry breaking potential. We address this aspect of the action in a later section. The essential idea here is to separate the scalar into gauge invariant and noninvariant pieces, then make a change of variables which removes the latter from the action. This transformation will also remove the gauge noninvariant components of the gauge field, seemingly for free. To do so, we follow \fa\ and \ma, where the field $\Phi$ is decomposed as follows (up to conventions), introducing the scalar $\rho\equiv |\Phi|$ and a matrix we will call $h$,
\be\label{h-def}
	\Phi = \rho  \left(\begin{array}{c} \Phi_1/\rho \\ \Phi_2/\rho \end{array}\right) = \rho  \left(\begin{array}{cc} \Phi_1/\rho  &-\bar{\Phi}_2/\rho \\ \Phi_2/\rho & \bar\Phi_1/\rho \end{array}\right)  \left(\begin{array}{c} 1 \\ 0\end{array}\right) \equiv \rho\, h[\Phi]  \left(\begin{array}{c} 1 \\ 0\end{array}\right)\;.
\ee
The question of whether this decomposition is allowed at $|\Phi|=0$ is actually crucial, and we return to it shortly. (The appearance of a fixed vector in (\ref{h-def}) suggests a relation to unitary gauge fixing, which we explain in the next section.) For now we observe that $\rho$ is gauge invariant, while the matrix $h$ is SU(2) valued, and under gauge transformations 
\be
	\Phi \to \Phi^U \equiv U^{-1}\Phi\;, \qquad A_\mu \to A_\mu^U\equiv U^{-1}A_\mu U + \frac{1}{g}U^{-1}\partial_\mu U\;,
\ee
it can be checked that $h$ transforms in the same way as the scalar, i.e.
\be\label{h-trans}
	h[\Phi] \rightarrow U^{-1}\, h[\Phi]\;.
\ee
This, as we will see, is the crucial property of $h$. Using this field-dependent matrix, we define from the gauge potential $A_\mu$ a new field $B_\mu$,
\be\label{B-def}
	B_\mu := h^{-1}A_\mu h + \frac{1}{g}h^{-1}\partial_\mu h\;.
\ee
Now, because of the transformation property of $h$, it follows that $B_\mu$ is {\it invariant} under local SU(2) transformations. If we rewrite the Lagrangian ({\ref{lag1}}) in terms of the new fields $\rho$, $h$ and $B_\mu$, we find
\be\label{lag2}
	\mathcal{L} = \partial_\mu\rho \partial^\mu\rho -\frac{1}{4}B^a_{\mu\nu}B_a^{\mu\nu}+\frac{g^2}{4}\rho^2 B^a_\mu B_a^\mu	+ \mathcal{L}_{ssb}(\rho)\;,
\ee
where $B_{\mu\nu}$ is the nonabelian field strength calculated from $B_\mu$ in the usual way, and all the dependence on $h$ has vanished. The remaining fields are all SU(2) invariant.  In this way, the gauge symmetry has been factored out of the action, but no gauge has been fixed. With the usual potential terms (\ref{ssb2}), the classical vacuum solutions following from (\ref{lag2}) are
\be\label{vacsolnew}
	B_\mu^a(x)=0\;,\qquad \rho(x) = \sqrt{\frac{-m^2}{\lambda}}\equiv v\;,
\ee
with $v$ being the scalar's v.e.v. Note, though, that these solutions are unique, in that there is a single value for the scalar in the minimum of the potential -- there is no U(1) of equivalent vacua. Expanding the scalar about this minimum, 
\be\label{expan1}
	\rho(x) = v + H(x)\;,
\ee
(\ref{lag2}) acquires mass terms and describes a massive vector and a single scalar, which are the expected degrees of freedom in the broken sector of the SU(2) Higgs theory. 

It is worth recalling here Elitzur's theorem, which states that it is impossible to spontaneously break a local symmetry \cite{Elitzur:1975im}. Essentially, the reason for this is that the functional integral averages over gauge orbits, so that even when the scalar's potential has a nontrivial minimum, the average of the scalar field will still be zero. Thus, the scalar acquires no v.e.v. and the gauge symmetry is preserved.  (See \cite{Fradkin:1978dv, Kajantie:1995kf, Gurtler:1997hr} for related investigations and \cite{Caudy:2007sf} for a recent review.) Starting from (\ref{lag1}), though, and proceeding to (\ref{lag2}), no gauge fixing is performed, nor are any properties of the scalar's potential employed. Performing only a change of variables, the result is a theory which is completely independent of the SU(2) symmetry. As there is no gauge symmetry, Elitzur's theorem has no consequence for the theory (\ref{lag2}). Taking the usual potential (\ref{ssb2}) we find the nontrivial vacuum solution (\ref{vacsolnew}), and expanding about this vacuum generates a mass for the field $B_\mu^a$. We will return to Elitzur's theorem later in the paper.

\subsection{The measure and conformal geometry}
In the interpretation of \cite{Chernodub:2008rz, Faddeev:2008qc}, we perform the above transformation but take $\mathcal{L}_{ssb}=0$. The classical vacuum solutions of (\ref{lag2}) are then degenerate, being,
\be
	B_\mu^a(x) =0\;, \qquad \rho^2(x) = \Lambda^2\;, 
\ee
for some $\Lambda$, the choice of which corresponds to a choice of vacuum. A reason for taking $\Lambda\not=0$, as required for mass generation, may be seen by quantising the theory. The functional measure for the new variables is
\be\label{measure1}
	\prod\limits_x\ \rho^2 \ud \rho^2 \ud B^a_\mu\;.
\ee
This is multiplied by an integral over the $h$ degrees of freedom, but since the Lagrangian is independent of $h$, this gives simply the volume of SU(2) and can be dropped. More interesting is the local factor of $\rho^2$ appearing in the measure. In \cite{Chernodub:2008rz, Faddeev:2008qc} this was interpreted as the conformal factor of a spacetime metric, $G_{\mu\nu}=\rho^2 \eta_{\mu\nu}$, since the Lagrangian (\ref{lag2}) is in just the right form to be rewritten
\be\label{lag3}
	\mathcal{L} = \sqrt{-G}\bigg( \frac{\mathcal{R}}{6} -\frac{1}{4}B^a_{\mu\nu}B_a^{\mu\nu}+ \frac{g^2}{4}B^a_\mu B_a^\mu\bigg)\;,
\ee
where all indices are now raised and lowered with respect to the metric $G$, and $\mathcal{R}$ is the Ricci scalar\footnote{See \cite{Chernodub:2008rz} for a discussion of the differences between the Euclidean and Minkowskian theories.} for $G$. This gives us a theory of a gauge boson living in a locally conformally flat spacetime. For phenomenological investigations of this idea see \cite{Foot:2008tz, Ryskin:2009kw}. Now, in order to maintain asymptotic flatness of the metric, we require $\rho^2(x) \to \Lambda^2$ asymptotically. In this way we can expand $\rho(x)=\Lambda + H(x)$, with $H(x)$ vanishing asymptotically, in analogy to (\ref{expan1}). The presence of $\Lambda$ generates the required mass terms in the action, in analogy to $v$ appearing in (\ref{vacsolnew}). Below, we will provide further physical motivation for the appearance of $\Lambda$.
\section{Relation to unitary gauge}\label{gauge-sect}
\subsection{The transformation}
Although we have only performed a change of variables, it is clear from (\ref{h-def}) and (\ref{B-def}) that there is a connection here to unitary gauge fixing. In this section we make this connection explicit. The purpose is to derive and understand a criteria, connected to scalar field's potential, for when the above transformation is valid.

We begin by noting that an implicit choice was made in the definition of $h$ in (\ref{h-def}) -- there are many other choices for the vector piece and ways to fill up the matrix. The only property of $h$ on which everything rests is that it should transform as in (\ref{h-trans}). The matrix $h$ is in fact just one example of a much more general object, called a `dressing', used to construct gauge invariant, physical fields in gauge theories \cite{Lavelle:1995ty}. The idea is that in order to describe physical particles, the gauge noninvariant Lagrangian fields must be combined into gauge invariant composites, or dressed, fields. The canonical example is Dirac's static electron $\Psi_s$ which is constructed from the Lagrangian fermion $\psi$ and the U(1) gauge field as \cite{Dirac:1955uv}
\be\label{dirac}
	\Psi_s = h^{-1}_s [A] \psi\equiv   \exp\bigg[ie \frac{\nabla.A}{\nabla^2}\bigg]\, \psi\;.
\ee
The exponential prefactor, or `dressing' $h^{-1}_s$, compensates for local gauge transformations of the fermion (but not for global transformations, so this electron does indeed carry a charge). The properties of dressed fields in both QED and QCD are discussed thoroughly in \cite{Lavelle:1995ty, Bagan:1999jf, Bagan:1999jk}. In particular, they allow for the construction of infrared finite asymptotic states and hence offer a route to  solving the infrared problems which plague gauge theories \cite{Lavelle:2005bt}.

Returning to our non--abelian theory, one could choose, in analogy to (\ref{dirac}), to construct a dressing using the gauge fields $A^a_\mu$.  This would give a transformation of (\ref{lag1}) to another set of variables which would, in complete analogy to $B^a_\mu$ and $\rho$, be gauge invariant. This is illustrated for the abelian Higgs model in \cite{Lavelle:1994rh}, but as it amounts to {\it directly} identifying the gauge invariant part of the gauge field, the same approach is much more difficult in  nonabelian theories. In fact, while it can be approached in perturbation theory, there is in general a nonperturbative obstruction, which we will encounter shortly. (In the present case we will also find a way around the problem.) However, \cite{Lavelle:1994rh} also shows that the interpretation of the invariant fields is more natural if one dresses using the scalars, i.e. with the $h$ given in (\ref{h-def}). Indeed, this choice makes the expected degrees of freedom of the broken sector manifest, just as we saw in (\ref{lag2}).

As shown in \cite{Lavelle:1995ty, Ilderton:2007qy}, dressings can be constructed using gauge fixing conditions. If we have a gauge condition $\chi[f]=0$, for some field $f$, then solving $\chi[f^h]=0$ gives the field dependent transformation $h\equiv h[f]$ which takes $f$ into the chosen gauge slice. From this one can show that $h^{-1}$ is a dressing. So, our matrix $h[\Phi]$ is indeed a gauge transformation, albeit a {\it field dependent} one: it is the transformation which takes an arbitrary field into unitary gauge. This is easily seen for the scalar field:
\be
	\Phi\to \Phi^h \equiv h^{-1}[\Phi] \Phi =\rho \left(\begin{array}{c} 1 \\ 0\end{array}\right)\;,
\ee
by definition, which is the single expected physical scalar of unitary gauge, while a short calculation gives the vector potential components, e.g., 
\begin{eqnarray*}
	B^3_\mu = &\frac{1}{|\Phi|^2}\bigg[ A^3_\mu(|\Phi_1|^2- |\Phi_2|^2) +\bar\Phi_1\Phi_2(A^1_\mu-i A^2_\mu)+ \bar\Phi_2\Phi_1(A^1_\mu+i A^2_\mu) \bigg] \\
	&+ \frac{i}{g|\Phi|^2}\bigg[ \bar\Phi_1 \partial_\mu \Phi_1 -  \Phi_1 \partial_\mu \bar\Phi_1 +  \bar\Phi_2 \partial_\mu \Phi_2 -  \Phi_2 \partial_\mu \bar\Phi_2\bigg]\;,
\end{eqnarray*}
and the others similarly.

\subsection{The measure as a Faddeev--Popov determinant}
We now offer a more standard interpretation of the functional measure in (\ref{measure1}). This will allow us to use familiar techniques to establish when the transformation discussed above holds nonperturbatively. Given the interpretation of the classical transformation, it is not too hard to see that the measure factor is just the Faddeev--Popov determinant for unitary gauge fixing. Let us briefly confirm this. Expanding the complex scalars as $\Phi_1= \varphi_1 + i\varphi_2$, $\Phi_2 = \varphi_3+ i\varphi_4$, unitary gauge eliminates the components $\varphi_2$, $\varphi_3$ and $\varphi_4$, so that $\rho \equiv |\Phi| \to |\varphi_1|$. Using the explicit form of the Pauli sigma matrices, one can calculate the variation of the gauge fixing conditions from their Poisson brackets with Gauss' law, e.g.,
\be\label{gauss}
	\{G^a\epsilon^a(x), \varphi_2(y)\} = (\epsilon^1\varphi_3 +\epsilon^2\varphi_4 + \epsilon^3\varphi_1) \delta(x-y)\;,
\ee
and check that the Faddeev--Popov determinant for this gauge condition is
\be\label{thedet}
	\left| \begin{array}{ccc}   \varphi_3 & \varphi_4 & \varphi_1 \\
						-\varphi_2 & -\varphi_1 & \varphi_4 \\
						\varphi_1 & -\varphi_2 & -\varphi_3 \end{array}   \right|_{\varphi_2=\varphi_3=\varphi_4=0} = |\varphi_1|^3\,.
\ee
The resulting measure on the remaining scalar is\footnote{We may trade $\varphi_1$ for $|\varphi_1|$, since the transformed action sees only $|\Phi|\to |\varphi_1|$.},
\be
	\prod\limits_x \varphi_1^3 \ud \varphi_1 \sim \prod\limits_x \varphi_1^2 \ud \varphi_1^2  \sim \prod\limits_x \rho^2 \ud \rho^2\,,
\ee
which is just the measure over the gauge invariant field $\rho$ given in (\ref{measure1}).

We are now ready to bring everything together. With some insight gained from the relations to unitary gauge, we can address when the above transformation is allowed, relate this to symmetry breaking and the condition of asymptotic flatness in the conformal interpretation.

\section{Physical degrees of freedom}\label{check-sect}
Recall that a good gauge fixing is identified by its Faddeev--Popov determinant being {\it non vanishing} for all field configurations. If it does vanish, we have a Gribov problem \cite{Singer:1978dk}. See \cite{Dell'Antonio:1991xt, Canfora:2008vt, Holdom:2009ws} for investigations of the distribution of Gribov copies, \cite{Heinzl:2007cp, vonSmekal:2007ns, Sorella:2009vt} for the connections to BRST invariance, and \cite{Dudal:2008sp, Bornyakov:2008yx, Greensite:2010hn} plus references therein for the effect of copies on the infra--red behaviour of ghost and gluon propagators, and the implications for the  Gribov--Zwaniger confinement conditions \cite{Zwanziger:1989mf}.

Now, since our functional measure is a Faddeev--Popov determinant, we may use the above condition to determine when our transformation is sensible quantum mechanically. It is  important to stress that if there is a Gribov problem, the new variables will {\it fail} to be gauge invariant nonperturbatively. Let us explain why this happens, and give a simple example.

Consider a U(1) gauge theory, with the {\it local} gauge transformation $U=\exp(ie\, k\cdot x)$ which sends $A_\mu \to A_\mu + k_\mu$. If a gauge field starts in Coulomb gauge, it returns to Coulomb gauge after this transformation, and so we {\it appear} to have Gribov copies. We will remove them in a moment, but first let us see how the copies affect dressed fields. Take Dirac's static electron (\ref{dirac}), and perform the above transformation. The dressed field behaves as
\be
	\Psi_s \to \exp(-ie k\cdot x)\, \Psi_s\;,
\ee
because $h^{-1}_s$ is {\it insensitive} to the transformation between copies. Our supposed electron then acquires a gauge dependence due to the Gribov ambiguity, and therefore does not describe a physical field. The resolution of this strange result is well known -- the fields and transformations in this example are not an allowed part of configuration space. The fields must fall off sufficiently fast at spatial infinity, and the gauge transformations must tend to the identity, or more generally the centre \cite{Lavelle:1995ty}, neither of which holds here, and so our copies are removed.

Contrastingly, one can construct in SU(2) explicit examples of allowed, finite energy configurations, their Gribov copies and the transformations between them \cite{Henyey:1978qd, vanBaal:1991zw, Grotowski:1999ay, Ilderton:2007qy}. These non--trivial copies  are explicitly nonperturbative, and, under transformations between them, the dressed fields indeed fail to be gauge invariant. 

This examples shows both the importance of boundary conditions when discussing copies, and the effect copies have on the gauge invariance of our transformed variables, which we have learnt are dressed fields. So, the Gribov ambiguity, whose presence is signalled by the vanishing of the Faddeev--Popov determinant, could introduce a nonperturbative gauge dependence into the supposedly invariant fields of (\ref{lag2}). In the case of SU(2)$\times$U(1), this would include the $Z$ and $W$ bosons, which are observable. There must, therefore, be a way to circumvent the Gribov problem in this case. We now want to show that this is the case, and how it is related to the scalar's potential. (For other approaches to treating the Gribov ambiguity see, e.g. \cite{Jackiw:1977ng, Reinhardt:2008pr}). 

We note that a gauge--covariant treatment of the Goldstone theorem shows that a spontaneously broken potential for the scalar (i.e. the potential (\ref{ssb2}) with $m^2<0$) will produce {\it gauge invariant} masses for the gauge fields in the Goldstone directions  \cite{O'Raifeartaigh:1990ht}. Thus, we can generate mass terms which are insensitive to Elitzur's averaging. (It is of course common to work in a particular gauge, and other aspects of the theory will reflect this choice.) To analyse when our Faddeev--Popov determinant vanishes, we therefore follow here the usual approach to the Higgs mechanism, beginning with (\ref{lag1}) and expanding the scalar around the minimum of its potential as so,
\be
	\varphi_1(x)\equiv v + H(x)\;,
\ee
in analogy to (\ref{expan1}). Now, finite energy considerations imply that the scalar fields of the theory vanish at infinity, see e.g.\ \cite{Ryder:1985wq}. Hence, we have $H(x)\to 0$. From the expressions (\ref{gauss}) we then have
\be
	\{G^a\epsilon^a(x), \varphi_2(y)\} =  (\epsilon^1\varphi_3 +\epsilon^2\varphi_4 + \epsilon^3\left(v + H(x)\right)  \delta(x-y)\;,
\ee
etc., and so the Faddeev--Popov determinant (\ref{thedet}) is nonvanishing for any allowed $H(x)$. This is because the only potentially difficult configuration, $H(x)=-v$, is forbidden by the boundary conditions. If, however, $v=0$ then the field
\be
	\varphi_1(x) \equiv H(x)=0
\ee
remains an allowed configuration and the determinant vanishes. (Going back to the original transformation, we noted that $|\Phi|=0$ was potentially problematic.) Therefore, when the scalar's potential has a trivial minimum, the Faddeev--Popov determinant, equivalently our functional measure, can vanish, which is a Gribov problem.

In the same fashion, suppose we take $\mathcal{L}_{ssb}=0$ and turn to the conformal interpretation, the functional determinant vanishes when there is a trivial asymptotic boundary condition $\rho(x)\to 0$. If, however, we impose asymptotic flatness via $\rho(x)\equiv \Lambda+H(x)$, with $\Lambda\not=0$ and $H(x)$ vanishing asymptotically, then the functional measure is nonvanishing and we can circumvent the Gribov problem. As we saw in Sect.~\ref{review}, $v$ and $\Lambda$ play the same role in mass generation, in the action. We have seen here that they also play the same role in avoiding the Gribov problem, and ensuring that the new variables are genuinely gauge invariant. 
\subsection{Extension to the Weinberg--Salam model}
The classical transformation to SU(2) invariant variables proceeds very similarly in the Weinberg--Salam model \fa. The matrix $h[\Phi]$ has precisely the same form as in (\ref{h-def}), and the same transformation property (\ref{h-trans}), which is the essential ingredient. The initial fields -- the doublet $\Phi$, SU(2) gauge field $A_\mu^a$ and U(1) gauge field $Y_\mu$ -- are transformed into
\be
	\rho\,,\quad Z_\mu\,,\quad W^+_\mu\,,\quad W^-_\mu\,,\quad A_\mu\;.
\ee
The U(1) of the original Lagrangian is not fixed by the transformation, and the resulting Lagrangian therefore describes a U(1) gauge theory, the photon $A_\mu$ being the remaining gauge non--invariant field. This abelian theory can be quantised by gauge fixing in the photon sector, as usual. We have the correct number $1 + 9 + 2=12$ degrees of freedom belonging to, respectively, the real scalar, three massive vector bosons and the photon $A_\mu$, with three SU(2) degrees of freedom factored out from the original Lagrangian. (The inclusion of fermions is straightforward \ma.)

As we have learnt, though, such a transformation only makes sense when conditions on the scalar field are such that we can circumvent the Gribov problem. Otherwise, Gribov copies introduce a nonperturbative gauge dependence into the new variables of the theory. Thus, we see that in the approach of \fa, the infra--red behaviour of the gravity sector has important consequences for the nonperturbative properties of the electroweak theory: the condition of asymptotic flatness is essential in order for the electroweak fields to be gauge invariant.

\subsection{Discussion}
In \ma\ it is suggested that the gauge invariant theory (\ref{lag2}), with the usual symmetry breaking potential (\ref{ssb2}), possesses two phases. If $m^2<0$ then the scalar can be expanded around the minimum of the potential, as above, generating the required mass terms. The other phase has $m^2>0$, and therefore no mass is generated. In this case, though, $\Phi\equiv 0$ is still part of configuration space, and while the classical Lagrangian (\ref{lag2}) appears to make sense, the transformation to it is not well defined because of the Gribov problem.

Indeed, the approach considered here will not work in all gauge theories, nor should it be expected to, as a straightforward counting of degrees of freedom can show \cite{Lunev:1994ty, Vlasov:1987vt}. Consider for example the $O(3)$ model which has three real scalar fields $\phi^a$ in the adjoint representation. After eliminating the $A^a_0$ multipliers, the theory describes a system with nine degrees of freedom -- there are twelve fields and three constraints coming from Gauss' law. The unitary gauge eliminates only two scalar degrees of freedom, and hence is not a complete gauge fixing.  It is, however, a good partial gauge fixing, provided the scalar has a v.e.v./asymptotic condition, i.e.\ we can expand $\phi^3(x) = v + H(x)$, with $\phi^1$, $\phi^2$ unchanged. For then we can check that the Faddeev--Popov determinant of these gauge conditions with the $G^1$ and $G^2$ components of Gauss' law does not vanish (hence, the set $\phi^1,\phi^2,G^1,G^2$ form a second class set of constraints). Direct calculation yields the nonvanishing Poisson brackets
\bea
	\{G^1(x),\phi^2(y)\} &= (v+H(x))\delta(x-y)\;,\\
	\{G^2(x),\phi^1(y)\} &= -(v+H(x))\delta(x-y)\;.
\eea
Thus, just as before, unitary gauge is a `good' gauge, but does not completely fix the gauge completely. In terms of the interpretations studied here, we could attempt to construct, from the scalar fields, a transformation to a gauge invariant set of variables. We know that this would be the field dependent transformation to $O(3)$ unitary gauge. While this exists, since unitary is a good partial gauge fixing here, the resulting variables would still transform under the unbroken part of the gauge group. They would not be gauge invariant.  Hence, there will be no dressing analogous to (\ref{h-def}) which can be used to define gauge invariant fields; equivalently, there does not exist a transformation to gauge invariant variables of the type considered above.

\section{Conclusions}\label{Concs}
The electroweak sector of the standard model may be expressed entirely in terms of fields which are explicitly SU(2) gauge invariant. We have shown that the transformation to these variables is the field--dependent gauge transformation to unitary gauge and, as such, can suffer a Gribov problem at the quantum level. This reintroduces, nonperturbatively, an unphysical gauge dependence into the new fields. We have shown, though, that this problem is circumvented precisely when the scalar field has a spontaneously broken potential. It is not surprising that the ability to identify physical variables in Higgs models has a subtle, but in this case explorable, dependence on the topology of the configuration space, as this is which really distinguishes non--abelian gauge theories from their abelian counterparts.

Note that it is not the breaking mechanism {\it per se} which is required to give a good transformation, but rather the existence of the nontrivial field minimum. Typically, of course, this comes about because of the potential terms. When the invariant scalar is interpreted as a conformal metric factor, though, the required `v.e.v.' of the field is generated by the condition of asymptotic flatness.  Again, this makes manifest the importance of topology on resolving the Gribov ambiguity: here it is the boundary conditions living on the `boundary' of spacetime which play a key role.

The claim in \cite{Masson:2010vx} that similar transformations to those given here can be applied to the whole standard model Lagrangian is a little suspect -- we know there are theories in which no such transformation exists. Additionally, when it does, the boundary conditions on the fields need to be carefully handled, see also \cite{Vlasov:1987vt}, as they have direct implication for the observability of the fields. Consider SU(3), where there is no symmetry breaking. In this case there are no physical descriptions of an observable asymptotic   gauge boson and  hence gluons are confined. 
\subsubsection*{Acknowledgements}

A. I. is supported by the European Research Council under Contract No. 204059-QPQV.

\end{document}